\documentclass[review]{elsarticle}

\usepackage{graphicx}
\usepackage{float}
\usepackage{amsmath}
\usepackage{subcaption}

\usepackage{lineno,hyperref}
\modulolinenumbers[5]
\journal{Atmospheric Research}

\begin{document}
\begin{frontmatter}
\title{Simulation of VHF signal generated by RREA}

\author[affil1,affil2]{Timur Khamitov\corref{mycorrespondingauthor}}
\cortext[mycorrespondingauthor]{Corresponding author}
\ead{timuruh@mail.ru}

\author[affil1,affil2]{Mikhail Zelenyi}
\author[affil3]{Ekaterina Svechnikova}
\author[affil1,affil2]{Alexander Nozik}

\address[affil1]{Moscow Institute of Physics and Technology - Institutsky lane 9, Dolgoprudny, Moscow region, 141700}
\address[affil2]{Institute for Nuclear Research of RAS - prospekt 60-letiya Oktyabrya 7a, Moscow 117312}
\address[affil3]{Institute of Applied Physics of RAS - 46 Ul’yanov str., 603950, Nizhny Novgorod, Russia}

\begin{abstract}
Relativistic runaway electron avalanches (RREAs) developing in electrified clouds produce gamma radiation which could be observed both on the ground surface and on near-Earth orbit. Radio emission of RREAs is a much less studied manifestation of the avalanche process. Meanwhile, VHF-radiation can be used as an additional source of information on conditions of the RREA development.

We present the method of calculating the VHF signal produced by RREAs using Monte-Carlo simulation with GEANT4. We show that VHF radiation produced by runaway electrons has a maximum at frequency range 0.1--1 MHz, the amplitude is below the background noise.
\end{abstract}


\end{frontmatter}


\section{Introduction}
Clouds in Earth's atmosphere are capable of producing gamma radiation, which could be detected from space as terrestrial gamma-ray flashes (TGFs)~\cite{Dwyer_2013_properties}, or measured by ground-based detectors as thunderstorm ground enhancements (TGEs)~\cite{Chilingarian_2011_natural_accelerator}.

The gamma radiation is produced mainly by bremsstrahlung of accelerated particles in relativistic runaway electron avalanches (RREAs) developed in the electric field of clouds~\cite{GUREVICH1992463}. 

For the so-called ``runaway'' electron the acceleration by electric field exceeds deceleration by the friction force in the atmosphere, leading to the gain of energy and increase of velocity~\cite{GUREVICH1992463}. Runaway electrons propagate through the air and collide with atoms. The runaway avalanche is possible for the electric field exceeding the ``critical'' value which was defined in~\cite{GUREVICH1992463}.

Runaway electrons propagating through the air and colliding with atoms emit electromagnetic waves, including gamma and radio. Gamma radiation of RREAs can be observed as TGFs and TGEs, spectra are thoroughly analysed in \cite{Briggs_2011, Marisaldi_2011, Mailyan_2016, Celestin_2014} and \cite{Mailyan_2013_TGE, Chilingarian_2019_extended_fluxes, Chilingarian_2019_origin} correspondingly. There are few studies devoted to the investigation of radio which could be produced by RREAs. The possible relation of radio pulses with two TGFs was discussed in~\cite{doi:10.1002/jgra.50188}, where characteristics of RREA produced by one seed electron were described. Meanwhile, it is known that the propagation of extensive air showers through the atmosphere is accompanied by radio emission~\protect\cite{LOFAR}. Therefore, data on the simultaneous detection of gamma and radio from electron avalanches in the atmosphere can be used as an additional source of information on atmospheric phenomena.

In this paper we investigate the radio-emission of relativistic particles in RREAs and discuss the possibilities of its detection in order to create a theoretical basis for future measurements. 

The theoretical analysis of radio-emission can help distinguish RREA models predicting similar emissions in gamma-range, as well as to separate RREA-related and other sources of observed radio-waves accompanying atmospheric phenomena. 

\section{Simulation details}
In order to investigate the gamma radiation of the avalanche we carry out a Monte-Carlo simulation using the GEANT4 simulation toolkit~\cite{geant4}. The electric field has a value higher than critical, which leads to avalanche multiplication of particles and formation of RREAs. 

Radio-emission of electrons is considered, while the contribution of positrons into the VHF signal is shown to be negligible. The minimal energy of a particle which behavior is simulated is 50~keV (the energy cut). This value is less than the minimum energy required for runaway processes (105~keV for 10~km altitude and 150~kV/m field) ~\cite{GUREVICH1992463}. Thus, the chosen energy cut makes it possible for us to consider the dynamics of all the particles significant for the RREA process. 

It should be noticed that the mean free path of a slow electron is around 1~mm ($l=\frac{M}{G*p}$, where M is the mass of $N_2$ molecule, p is atmosphere density and G is a cross section of collision for $N_2$~\cite{NIST}), while electron gains only hundreds of eV per 1~mm from field's acceleration.

Particles with energy below the cut energy have a negligible probability to gain the energy above critical and become runaway. Thus, the chosen value of the energy cut enables us to model all the processes significant for the development of a RREA. 

An important feature of the simulation with GEANT4 is that the modeled media (air in our case) affect moving particles but can not be influenced by these particles. In other words, the media defines the friction force and cross-sections of processes of particle-media interactions. The processes of charge accumulation in the media and ion drift are not calculated. Also, the modeling results obtained using GEANT4 are known to depend on the physics list~\cite{Skeltved_2014}. We used ``G4EmStandardPhysics'' and checked that the difference from the results obtained with other physics lists (Penelope and Livermore models) is smaller than the difference between different runs within one physics list.

The simulated region is a 1~km height cylinder filled with air, the electric field is directed along the axis of the cylinder. Two cases with different conditions were analysed: air density corresponding to an altitude of 5~km with an applied electric field of 200~kV/m and air density corresponding to 10~km with the electric field of 150~kV/m.
Each avalanche is caused by one electron with an energy of 3 MeV.

We use only one initial electron as a seed because two or more energetic particles of secondary cosmic rays are very unlikely to participate in the development of one avalanche.
The flux of cosmic electrons at the altitude 10~km is estimated by EXPACS program~\cite{EXPACS} as 0.368~$cm^{-2}s^{-1}$ for electrons with energy bigger than 450 keV (Gurevich's critical energy required for runaway~\cite{GUREVICH1992463}), i.e. 0.368 electrons per 100 square meters per microsecond which is less than one particle during the lifetime of each avalanche pulse. A similar estimation can be obtained using the data on cosmic rays from~\cite{flux}.

\begin{figure}[H]
\centering
\includegraphics{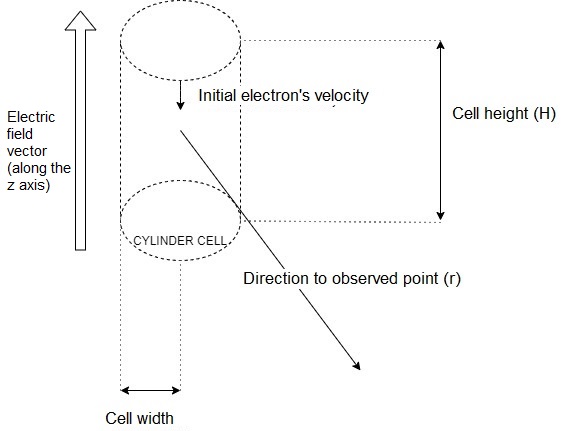}
\caption{The modeled region is filled with air with density equal to 5 or 10~km altitude for different simulations. The electric field is directed along the z-axis and exists only inside the cylinder. The height of the cylinder is 1~km, the diameter is 5~km.}
\end{figure}

GEANT4 provides us by coordinates and time of discrete steps of tracks.That gives a 
possibility to calculate velocity and acceleration of runaway electrons and use it further for calculation according to the expression~\ref{LNP} derived from Liénard–Wiechert potentials~\cite{landau}:
 \begin{equation}\label{LNP}
              E(\boldsymbol{r} ,t) = \frac1{4\pi \varepsilon_0} ( 
           \frac{q ( \boldsymbol{n}-\boldsymbol{\beta})}{\gamma^2(1-\boldsymbol{n\beta})^3|\boldsymbol{r}-\boldsymbol{r_s}|^2}+\frac{q\boldsymbol{n}\times((\boldsymbol{n}-\boldsymbol{\beta})\times\boldsymbol{\dot{\beta}})}{c(1-\boldsymbol{n\beta})^3 |\boldsymbol{r}-\boldsymbol{r_s}|} )|_{t_r} 
   \end{equation}
   , where
   $\boldsymbol{r}$ is particle position vector,
    $\boldsymbol{r_s}$ is vector of direction to observation point, 
     $\boldsymbol{n}=\frac{\boldsymbol{r}-\boldsymbol{r_s}}{|\boldsymbol{r}-\boldsymbol{r_s}|}$ is direction from particle to observer, 
      $\boldsymbol{\beta}(t)=\frac{\boldsymbol{v_s}(t)}{c}$ is relativistic velocity,
      $\gamma(t)=\frac{1}{\sqrt[2][1-|\boldsymbol{\beta}(t)|^2]}$ - the Lorentz factor, 
      $\boldsymbol{\dot{\beta}}(t)$ stands for relativistic acceleration, 
      $\ t $ is time,
      $\ t_r=t-\frac{|\boldsymbol{r}-\boldsymbol{r_s}|}{c} $ is retarded time,
     $\boldsymbol{E_{av}}=\frac{\sum\limits_{i}^{N}\sum\limits_{t}^{M_i} \boldsymbol{E_{ji}}}{N} $is averaged electric field, 
     $\boldsymbol{E_{ji}}$ is electric field obtained from particle of type $j$ in simulation $i$, N is number of simulations and $M_i$ is number of particles in simulation $i$.

\section{Results}
\begin{figure}[ht!]
    \begin{subfigure}[b]{0.5\textwidth}
    \includegraphics[width=0.95\textwidth]{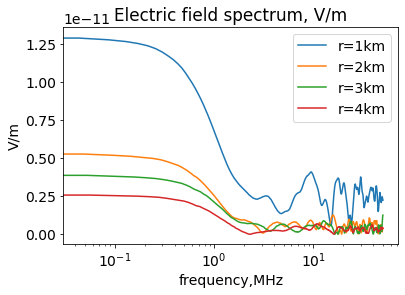} 
    \end{subfigure}
	\begin{subfigure}[b]{0.5\textwidth}    \includegraphics[width=0.95\textwidth]{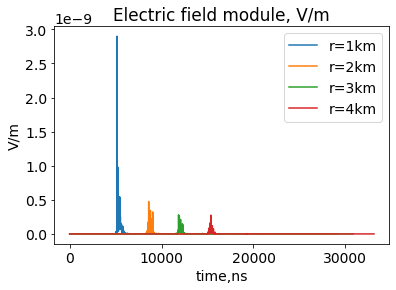}
    \end{subfigure}
    \caption{
   The time structure of the signal and the spectrum of VHF radiation created by the RREA, on the vertical axis of the cylindrical modeled region. Calculation is carried out for air density corresponding to 5~km altitude and electric field strength~200~kV/m. $r$ is a distance from the center of the cylindrical modeled region to the observation point.}
    \label{Fig: 5km}
\end{figure}

\begin{figure}[H]
    \begin{subfigure}[b]{0.5\textwidth}
    \includegraphics[width=0.95\textwidth]{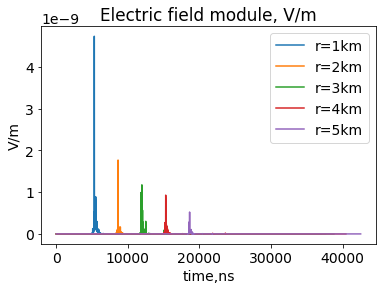} 
    \end{subfigure}
	\begin{subfigure}[b]{0.5\textwidth}    \includegraphics[width=0.95\textwidth]{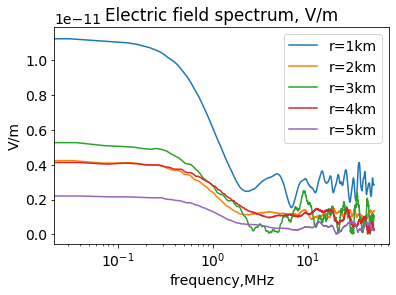}
    \end{subfigure}
    \caption{
    The time structure of the signal and the spectrum: 10 km height air, 150 kv/m field.
       }\label{Fig: 10km}
\end{figure}

Figures~\ref{Fig: 5km} and~\ref{Fig: 10km} present the VHF signal of one avalanche averaged over 10000 runs of the simulation. It should be noticed that the development of RREA and the resulting radio signal are very different in different runs, so the averaged picture does not represent most of the signals.

A variety of cases of avalanche development is caused by the chaotic nature of processes with the seed electron and the exponential increase in the number of particles~\cite{GUREVICH1992463}. Thus, random interactions of the seed electron very strongly affect the development of the RREA. In particular, the number of particles in the avalanche varies from one to about~$10^5$.

\begin{figure}[H]
    \begin{subfigure}[b]{0.5\textwidth}
    \includegraphics[width=0.95\textwidth]{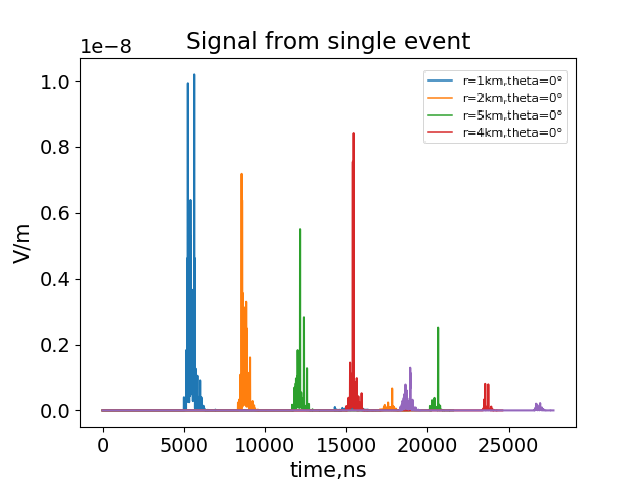} 
    \end{subfigure}
	\begin{subfigure}[b]{0.5\textwidth}    \includegraphics[width=0.95\textwidth]{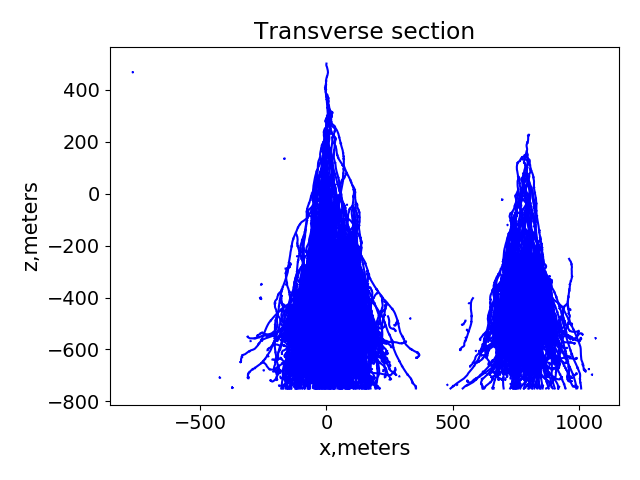}
    \end{subfigure}        
    \caption{
    Vertical cross-section of trajectories of particles of the RREA and VHF-signal of the RREA. The two ``branches'' are generated by gamma-quanta emitted by the avalanche. The interference between the two branches lead to local maximum at 5~km under the modeled region and postponed signal at 4~km observed point due to non-vertical movement of particles.
       }\label{Fig: interf}
\end{figure}

The amplitude of the signal is below the typical background level in the atmosphere~\cite{background}, thus the radio-signal of one RREA hardly can be registered.
The dependence of the amplitude of the VHF signal on the horizontal coordinate demonstrates oscillations which can be significant for future measurement applications. 
Initially, electrons of the avalanche electrons are localized in space, forming a``branch''. Particles of this branch can create new branches, spatially separated from one another and from the initial one.
We assume that oscillations are caused by interference of radiation emitted by different branches of the RREA - spatially separated parts of the avalanche shown at~Fig.\ref{Fig: interf}.

The number of runaway electrons is less than the number of all the other (``slow'') electrons in the RREA by about~$10^4$~times (according to~\cite{Dwyer_Babich_2011} and our calculations), that is why radio-signal produced by slow electrons can be much stronger than that of runaway electrons. The period of the oscillation of the signal can be estimated as the time required for the electron to reach the bottom of the cell - it gives around 0.24 MHz. Here we suppose the average speed $v$=$0.8*c$ , where c is the speed of light.

\section{Discussion}

Our analysis provides an estimation of VHF signal, which is a part of the radiation of RREA produced predominantly by runaway electrons. The spectrum of the RREA without consideration of impacts of different groups of particles presented in~\cite{doi:10.1002/jgra.50188} begins below 100 MHz and has a maximum at 100 kHz - 1 MHz frequency. The amplitude reported in~\cite{doi:10.1002/jgra.50188} is much greater than the value calculated for runaway electrons in the present paper, because the number of slow electrons is $10^4$ times bigger than the number of runaway electrons. Thus, the signal produced by slow electrons is much stronger than that of runaway electrons.  
We aimed to estimate the VHF signal of RREAs, which is mainly produced by runaway electrons.

\begin{figure}[H]
    \centering
       \includegraphics[scale=0.7]{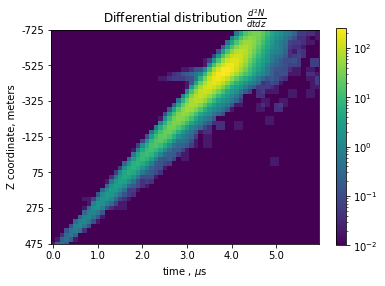}
    \caption{
    Number of electrons in (t, t+dt)$\times$(z, z+dz) in the RREA developed in the electric field 150~kv/m at the altitude 10~km, where $dt=100 ns$ and $dz=25 m$.
       }\label{Fig: NumberDiff}
\end{figure}

The result of the modeling is in agreement with an estimation obtained on the basis of the Formula~\ref{RakovFormula} introduced in~\cite{doi:10.1029/2010JD014235}, describing the electric field generated by electrical currents. We compared the result of our simulation with the electric field found with the Formula~~\ref{RakovFormula} for currents calculated in the simulation. Only the second term in the Formula~\ref{RakovFormula} is significant for estimation of the radiation of RREAs because the first term decays with distance rapidly, the third term turns into zero for the observation point located directly under the center of the region with the current. While calculating the current we took into account only vertical components of velocities and multiplied by the number of particles shown at~Fig.\ref{Fig: NumberDiff} in order to average high-frequency oscillations and corresponding thin peaks shown at~Fig.\ref{Fig: 5km}-\ref{Fig: 10km}.

\begin{figure}[H]
    \centering
    \includegraphics[scale=.65]{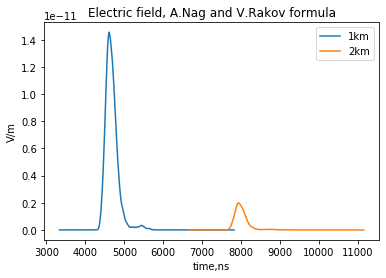}
      \caption{
    Signal dependence on time calculated with the Formula \ref{RakovFormula}
    }\label{Fig: RakovFigure}
\end{figure}

\begin{multline}\label{RakovFormula}
    E_z(r,t)=\frac{1}{2\pi \epsilon_0}
    \int_{h_1}^{h_2}   [ dz \frac{2z^2-r^2}{R^5 (z)} \int_{t_b (z)}^{t} i(z,\tau-\frac{R(z)}{c})d\tau \\
    +\frac{2z^2-r^2}{cR^4 (z)}i(z,\tau-\frac{R(z)}{c})dz- \frac{r^2}{c^2R^3 (z)}\frac{di(z,\tau-\frac{R(z)}{c})dz}{dt}dz ]
\end{multline}
$ R (z) $ is the distance from the point with vertical coordinate $z$ to the observation point, $ h_1 $ and $ h_2 $ are the altitudes of the bottom and the top of the region with current correspondingly. 
$t_b (z) $ is the moment the radiation reaches the point with vertical coordinate $z$, r is the horizontal distance to the observation point, 
$i(z,\tau-\frac{R (z)} {c})$ is the current strength at the moment when the radiation reaches the observation point. 

Fig.\ref{Fig: NumberDiff} shows the differential distribution of the number of particles in a RREA developed in the 150 kV/m field at altitude 10~km, modeled using GEANT4. It is noticeable that the avalanche starts to decay at the bottom of the modeled region ($z=-500 m$) at the time required for electrons to reach the bottom. The modeled dynamics of electrons density was used for the calculation of the current distribution.

\section{Conclusions}
Within the framework of Gurevich's model of the development of the RREA, we showed that the emission spectrum of runaway electrons of the RREA almost entirely fits into the interval 0.01 MHz--100 MHz having the maximum at 0.1--1 MHz. The shape of the spectrum depends on the size of the region of the strong electric field because this size limits the length of RREAs. Non-relativistic particles (slow electrons and ions) can produce a strong radio-signal, which should be considered in further research.

The dependence of the VHF signal on the horizontal coordinate can have a sequence of local maximums caused by interfering radiation of different parts of the RREA. The amplitude of the VHF signal is below the natural background. The oscillation of the signal over the horizontal coordinate requires further study and can provide information on the structure of the avalanche. 

The authors express their gratitude to Maxim Dolgonosov and Lev Zelenyi for fruitful discussions.

\bibliography{refs}

\end{document}